# Enhancing Recommender System Performance by Histogram Equalization

Hao Wang
CEO Office
Ratidar Technologies LLC
Beijing, China
Haow85@live.com

*Abstract*—Recommender system has been researched for decades with millions of different versions of algorithms created in the industry. In spite of the huge amount of work spent on the field, there are many basic questions to be answered in the field. The most fundamental question to be answered is the accuracy problem, and in recent years, fairness becomes the new buzz word for researchers. In this paper, we borrow an idea from image processing, namely, histogram equalization. As a preprocessing step to recommender system algorithms, histogram equalization could enhance both the accuracy and fairness metrics of the recommender system algorithms. In the experiment section, we prove that our new approach could improve vanila algorithms by a large margin in accuracy metric and stay competitive on fairness metrics.

I. INTRODUCTION

Consumers love recommendations, and this is why marketing has become a lucrative industry. However, as human beings entered the internet age, modern information technologies have substituted a lion share of the marketing industry with computational advertising and recommendation tools. Imagine the recommendation team in you company comprises of around 30 people. It would cost millions of USD per year for the salaries. However, for a huge company like Amazon.com, whose online sales volume has been increased by 30% by recommendation team, the team cost is only a tiny fraction of the total revenue generated by the technologies. This is the reason why both the industry and the academia are keen in the research and development of the technology.

Increasing the accuracy of the recommender systems is the key goal of the industrial practitioners. In order to lower the churn rate and increase the traffic volume, industrial practitioners update their algorithmic work on a frequency basis of weeks. Since the field started in 1990's, there have been millions or tens of millions of different versions of algorithmic work by 2023. However, since most industrial work is only micro-innovation or engineering oriented work, disruptive innovation is still rarely seen compared with the huge amount of work contributed by programmers globally.

In early days of recommender systems, people mainly focused on the RMSE or MAE metric of the system, namely, they want to make the prediction score of users' future preferences close to the true preference of customers. Later, researchers started to explore other metrics such as NDCG, AUC, Precision@K, etc. Instead of minimizing the gap between estimated preference score and true score, people started to optimize the ranking order of the recommendation results. New technologies such as Deep Learning started to swarm the field around 2016 [1][2], and since then, the industrial models have become more and more complicated.

One intrinsic problem associated with recommender systems is fairness. Different types of unfairness problems such as popularity bias, selection bias, exposure bias, demographic unfairness have plagued the field since the very beginning. Researchers have applied regularization and other techniques to solve the problems, in the hope to create products that are not only precise but also humane.

Another field that has experienced dramatic change in the past 10 years is image processing. The key driving force behind the image processing revolution is deep learning. Instead of using shallow features such as SIFT, deep learning models are capable of solving highly complicated tasks with accuracy unattainable in the past experiments. Industrial missions such as autonomous driving, safety helmet recognition, equipment fault detection, etc. have experienced a revolution thanks to deep learning.

In this paper, a bridge between the field of recommender system and image processing are created to improve performance of recommendation engines. The old concept of histogram equalization is introduced into the field of recommender system, and hybrid models are created to produce algorithmic models that are both accurate and fair.

## II. RELATED WORK

Recommender systems focuses on improving recommendation results to increase sales and traffic volumes. Early models such as collaborative filtering and matrix factorization have dominated the field for decades. Classic algorithms include item-based collaborative filtering [3], SVD++ [4], SVDFeature [5], ALS [6], among many different variants. Later inventions include learning to rank approaches such as BPR [7] and CLiMF [8]. As time goes by, the technical paradigms of recommender systems have evolved into deep learning approaches such as DeepFM [9], Wide & Deep [10], AutoInt [11], etc.

There has been research on intrinsic issues associated with recommender systems such as popularity bias [12][13], selection bias [14][15], exposure bias [16][17], etc. Important technical benchmark in this field include learning to rank [18][19] and matrix factorization [20][21]. Another issue other than fairness is the cold-start problem. Cold-start problem is a problem encountered by nearly every recommender system builder. Researchers have utilized transfer learning [22][23] and zeroshot learning [24][25] to solve the problem.

Image processing is a field that has experienced dramatic change in recent years. Since the advent of AlexNet [26], deep learning models [27][28][29] have flooded the field. However, old concepts such as histogram equalization [30] is also important in the field, as will be discussed in the following sections of the paper.

## III. HISTOGRAM EQUALIZATION

Image processing is a field rich in algorithmic innovation. One of the earliest concepts in the field is histogram equalization. The motivation behind histogram equalization is to transform the color histogram of the images into a more equalized one so that the human perception of the images would be enhanced. In formal terms, assume the probability density function of image pixel values follow distribution P. Then after histogram equalization, pixel values of the new image take values according to the following formula :

$$S(K) = \frac{1}{L-1} \sum_{k=1}^{K} P(r_k)$$

, where $r_k$ represents the pixel values of the image before transformation, whose maximum value is L-1, and k denotes the k-th level of the greyscale values of the image. P is the frequency ratio of $r_k$.

The basic idea behind histogram equalization is to transform the initial image into a new image whose greyscale values are as equally distributed as possible. This idea could be easily extended into the field of recommender systems - one of the major reasons of the popularity bias issue is the popularity bias problem in the input structure. By applying the histogram equalization algorithm to the input user item rating values of the recommender system, the goal to smooth out the head of the long tail and level up the tail of the distribution is achieved. Formally :

$$R(K) = R_{max} \sum_{k=1}^{K} P(R_k)$$

R denotes the user item rating values of the user item rating matrix.

By applying histogram equalization algorithm to the user item rating matrix values, transformation from the original highly skewed dataset to a more uniformly distributed data structure is implemented. A natural question arises : How should the performance of the algorithm be evaluated in the end ? Should ranking metrics such as Precision@K be used rather than accuracy metrics such as MAE and RMSE ? A bold step is taken to preserve the classic accuracy metrics for evaluation in our experiments.

The formal procedure of the algorithm goes as follows : 1. Apply histogram equalization to user item rating matrix 2. Apply recommender system approach to the transformed user item rating matrix 3. Evaluate results using MAE (or other metrics, in compliance of the practical needs).

Both the fields of recommender systems and image processing have evolved for decades, and it's rare to see a paper that builds the connections between the 2 areas. To our humble knowledge, this paper is the first of its kind to create connections between the 2 fields other than deep learning models. Although the development of the theory might seem simple and trivial, it is the first step towards a more complicated unified understanding of different computer science domains.

The input data structures to recommender systems are essentially 2-D data arrays, and the input data structures to image processing algorithms are also 2-D data arrays. Recommender system technologies such as matrix factorization can be imported into the field of image processing for tasks such as image compression and restoration directly. Other connections between the 2 fields are still left for exploration for researchers. This is the future direction of our research and enough time and energy will be spent on the topic.

## IV. EXPERIMENTS

Our algorithm is tested with other algorithms without histogram equalization on MovieLens 1 Million Dataset [31] and LDOS- CoMoDa Dataset [32]. The algorithms chosen for the experiments are vanilla matrix factorization and KL-Mat [33]. MovieLens 1 Million Dataset comprises of 6040 users and 3706 movies, while LDOS-CoMoDa Dataset is composed of 121 users and

1232 movies. The results are shown in the following figures :

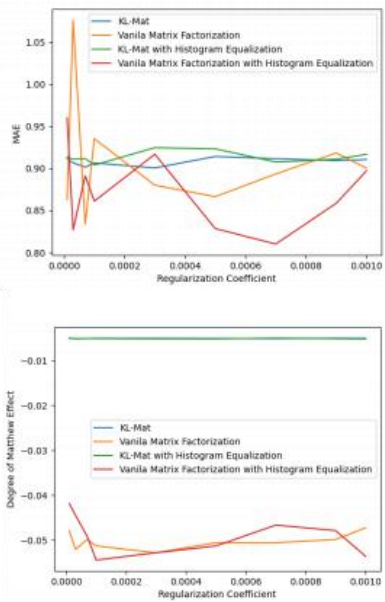

Fig.1 MovieLens 1 Million Dataset

Fig.1 demonstrates the comparison results among different algorithms on the MovieLens 1 Million Dataset. By the figures, Histogram Equalization enhances vanilla matrix factorization accuracy performance by a large margin, and is on par with the vanilla algorithms on fairness metrics.

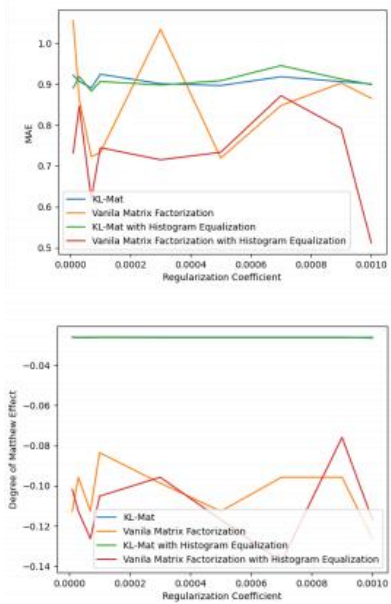

Fig.2 LDOS-CoMoDa Dataset

Fig.2 demonstrates the comparison results among different algorithms on the LDOS-CoMoDa Dataset. By the figures, Histogram Equalization enhances vanilla matrix factorization accuracy performance by a large margin, and is on par with the vanilla algorithms on fairness metrics.

## V. CONCLUSION

In this paper, a new technique for recommender systems that relies on a classic image processing idea, namely histogram equalization is introduced. The old concept has instilled new blood into the field of machine learning, and produces very competitive results, as demonstrated in the experiment section.

In future work, we would like to explore other image processing ideas such as image matching, etc. and their applications in the machine learning fields. Our long term goal of our research in this direction is to create a unified theory between recommender systems and image processing. We believe cross-domain research can lead to results unimaginable by other researchers.